\documentstyle[aps,pre,epsf,floats,amsmath,amssymb]{revtex}

\def\headline#1{\noindent{\bf #1}\\ }

\begin{document}
\parskip 2mm
\twocolumn[\hsize\textwidth\columnwidth\hsize\csname@twocolumnfalse%
\endcsname
\title{First-order transitions in fluctuating \\
1+1-dimensional nonequilibrium systems}
\author{Haye Hinrichsen\\[2mm]}
\address{Theoretische Physik, Fachbereich 10,\\
     Gerhard-Mercator-Universit{\"a}t Duisburg,
     47048 Duisburg, Germany}
\date{June 13, 2000}
\maketitle

\begin{abstract}
We demonstrate that first-order phase transitions in 1+1-dimensional 
nonequilibrium systems with fluctuating ordered phases are impossible,
provided that there are no additional conservation laws, 
long-range interactions, macroscopic currents, or 
special boundary conditions. Since minority islands in the
ordered phase of such systems can only shrink by short-range interactions,
it is impossible to stabilize a fluctuating ordered state. The
apparent first-order behavior turns out to be a transient
phenomenon, crossing over to a continuous transition after very
long time. As examples we consider the triplet creation model
$3X\rightarrow 4X$, $X\rightarrow \O$, the annihilation/fission
process $2X\rightarrow 3X$, $2X\rightarrow \O$, as well as
spreading on a diffusing background $X$+$Y\rightarrow 2Y$,
$Y\rightarrow X$.
\end{abstract}

]
%

\section{Introduction}

In equilibrium statistical mechanics it is well known that systems
undergoing a first-order phase transition in high space dimensions
$d$ may exhibit a second-order transition below their upper
critical dimension. The reason is that in low dimensions
fluctuations become more important and may destabilize the ordered
phase. A very similar situation emerges in the case of
nonequilibrium phase transitions~\cite{MarroDickman99}. In this
context the question arises under which conditions first-order
phase transitions can be observed in one spatial dimension. The
purpose of the paper is to point out that various 1+1-dimensional
nonequilibrium models, which were believed to exhibit a
first-order transition, cross over to a continuous transition
after very long time. 

Dynamic random processes with first-order phase transitions are
usually characterized by several stable ordered phases. For
example, in the subcritical regime $0<T<T_c$ of the
two-dimensional kinetic Ising model there are two stable
magnetized states. To ensure their stability, the Ising model
provides a robust mechanism eliminating islands of the minority
phase generated by thermal fluctuations. This mechanism relies on
the fact that the boundary of an island costs energy, leading to
an effective {\it surface tension}. Attempting to minimize its energy,
the island is subjected to an attracting `force' and begins to
shrink. It is important to note that this long-range force
decreases {\em algebraically} as $1/r$ with the typical radius $r$
of the island so that thermal fluctuations of any size are safely
eliminated. Because of the $Z_2$-symmetry under spin reversal,
both ordered phases are equally attracting. Thus, starting from a
disordered state with zero magnetization, we observe coarsening 
patterns of ordered domains. However, if an external field is applied,
one type of minority islands becomes unstable above a certain
critical size. Since there is a finite probability to generate
such islands by fluctuations, one of the two ordered phases 
eventually takes over, i.e., the system undergoes a first-order 
phase transition.

\begin{figure}
\epsfxsize=75mm
\centerline{\epsffile{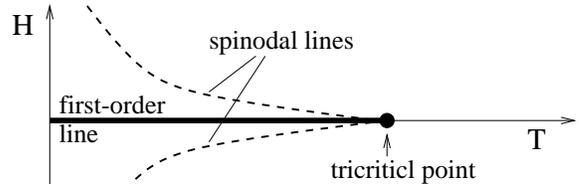}}
\caption{
\label{FIGTOOM}
Schematic  phase diagram of Toom's north-east-center voting model.
}
\end{figure}

Turning to nonequilibrium systems, the mechanism for the
elimination of minority islands may be even more robust. An
interesting example is Toom's two-dimensional north-east-center
voting  model~\cite{Toom80,BennettGrinstein85}. Again there are
two stable ordered phases. In contrast to the kinetic Ising model,
where straight domain walls perform an unbiased diffusive motion,
the interfaces in Toom's model propagate in a preferred direction,
wherefore the process is out of equilibrium. As a key property of
the model, this propagation velocity depends on the orientation of
the domain walls. Using this anisotropy, the dynamic rules of the
model are designed in a way that minority islands quickly assume a
triangular shape and begin to shrink with constant velocity. Thus,
the effective `force' by which an island shrinks is
independent of~$r$, leading to much more stable phases as in the
standard kinetic Ising model. Consequently, the ordered phases
remain stable even if an oppositely oriented external field $H$
is applied. In order to flip the whole system, the intensity of
the field has to exceed a certain critical threshold. The
corresponding phase stability boundaries (also called spinodal
lines) are sketched in Fig.~\ref{FIGTOOM}. Between these lines,
the two ordered states are both thermodynamically stable, i.e.,
they coexist in a whole region of the parameter space. Crossing
the coexistence region by varying the external field, the
magnetization of the system follows in a hysteresis loop. Similar
phenomena can be observed in certain models for nonequilibrium
wetting~\cite{HLMP00}.

Obviously, both mechanisms for the elimination of minority islands
-- surface tension and anisotropic propagation velocities -- can
only be implemented in at least two spatial dimensions. Therefore,
it is interesting to investigate the question under which
conditions first-order phase transitions can be observed in one
spatial dimension. For example, the 1+1-dimensional
Ziff-Gulari-Barshad model~\cite{ZGB86} for heterogeneous catalysis
is known to exhibit a first-order phase transition which relies on
the interplay of three different kinds of particles. Similarly, a
recently introduced model for phase separation on a
ring~\cite{EKKM98} uses three different species of particles.
Another example is the so-called bridge model~\cite{GLEMSS95} for
bidirectional traffic on a single lane, where special boundary
conditions induce a discontinuous transition in the currents. Even
more subtle is the mechanism in the two-species model introduced
in a recent paper by Oerding {\it et al.}, where a first-order
phase transition is induced by fluctuations~\cite{OWLH99}.
Therefore, in attempting to comprehend the full range of
first-order phase transition under nonequilibrium conditions, it
would be interesting to seek for the simplest 1+1-dimensional
model which exhibits a discontinuous transition. By `simple' we
mean that such a model should involve only one species of
particles evolving by local dynamic rules without macroscopic
currents, conservation laws, and unconventional symmetries.
Moreover, the choice of the boundary conditions should be
irrelevant.

The prototype of such a dynamic process is the 
one-dimensional Glauber-Ising model
at zero temperature in a magnetic field. This model, which is also
referred to as {\em compact directed percolation} (CDP), can be
defined as follows. Sites of a one-dimensional lattice can be in
two different states $\uparrow$ and $\downarrow$. For each update
a pair of adjacent sites is randomly selected. If the two spins
are in opposite states, the are aligned with the probabilities $p$
and $1-p$:
\begin{equation}
\uparrow\downarrow,\downarrow\uparrow \;\;
\stackrel{p}{\longrightarrow} \;\;\uparrow\uparrow\,,\qquad
\uparrow\downarrow,\downarrow\uparrow \; \;
\stackrel{1-p}{\longrightarrow} \;\;\downarrow\downarrow \,.
\end{equation}
Obviously, the parameter $H=p-1/2$ plays the role of an external
field. Since there is no temperature, the model has two {\it
absorbing configurations}, namely the fully magnetized states
$A=...\downarrow\downarrow\downarrow...$ and
$B=...\uparrow\uparrow\uparrow...$ . Once the system has reached
one of the two absorbing configurations, it is trapped and
will remain there forever.

In the off-critical case $H \neq 0$, one of the two absorbing
states is stable while the other one is unstable against small
perturbations. For example, for $H<0$, $B$-islands tend to grow
while $A$-islands shrink. Thus, starting with random initial
conditions, the system approaches the thermodynamically stable
state $B$ in an exponentially short time. In the vicinity of the
phase transition this time scale diverges as $|H|^{-1}$.
Right at the critical point, the two absorbing states of the
Glauber model are  only {\em marginally} stable against
perturbations. For example, by flipping a single spin in a fully
magnetized domain, we create a pair of kinks. These kinks perform
an unbiased random walk until they annihilate one another. Thus,
minority islands do not shrink by virtue of an attracting force,
rather they are eliminated solely because of the fact that random
walkers in one dimension always return to their origin.
Consequently, the lifetimes $\tau$ of minority islands are finite
and distributed algebraically as $P(\tau) \sim \tau^{-1/2}$. It is
important to note that this mechanism allows islands to reach a
macroscopic size of the order $\sqrt{\tau}$ during their temporal
evolution. Therefore, averaging over many independent samples, the
mean size of such minority islands approaches a constant value. As
a consequence, the ordered phase is only marginally stable against
perturbations. In fact, introducing a small rate for spontaneous
spin flips, the first-order transition in the Glauber-Ising model
is lost. The same happens in certain nonequilibrium Ising models
with two absorbing states subjected to an external
field~\cite{MenyhardOdor95,MenyhardOdor98}.

Is it possible to observe first-order transitions in {\em
fluctuating} 1+1-dimensional two-state systems? Obviously, such a
model requires a much more robust mechanism for the elimination of
minority islands. A simple random walk of a pair of kinks is not
sufficient, rather there has to be an attracting force which
prevents small island from growing. It seems that such a mechanism
is difficult to implement. In fact, various models, which have
been suggested in the past, display a discontinuous transition
only in $d \geq 2$
dimensions~\cite{RBC89,Jensen91,Albano92,OBS93,MenyhardOdor95,OdorSzolnoki96}.

A frequently cited exception is the 1+1-dimensional triplet
creation process introduced by Dickman and
Tom\'e~\cite{DickmanTome91}:
\begin{equation}
\label{DickmanTomeProcess}
3X\rightarrow 4X\,, \quad X\rightarrow \O \,.
\end{equation}
In this model the high-density phase is not strictly absorbing,
rather islands of unoccupied sites are spontaneously created in
the bulk so that the active state fluctuates. In numerical
simulations it was observed that above a certain tricritical point
the second-order phase transition line splits up into two spinodal
lines, where the transition becomes first order. Moreover, the
order parameter seemed to follow a hysteresis loop when the
parameter for offspring production was varied. Apparently, the
highly nonlinear particle creation process $3X\rightarrow 4X$ is
able to stabilize the high-density phase, eliminating islands of
the minority phase.
Similarly, Carlon {\it et al.}~\cite{CHS99} reported a first
order transition in the 1+1-dimensional annihilation/fission
process on a lattice~\cite{HowardTauber97}
\begin{equation}
\label{AFProcess}
2X\rightarrow 3X\,, \quad 2X\rightarrow \O\,.
\end{equation}
For low values of the diffusion constant 
they found a continuous transitions which
becomes first order for high diffusion rates above a certain
tricritical point. 

In the present work we argue that this type of first-order
behavior in 1+1-dimensional models is only a transient phenomenon.
Although mean field approximations predict a discontinuous
behavior, the transition crosses over to a continuous transition
for any value of the parameters (i.e., for any value of the
diffusion constant in the examples mentioned before).

The paper is organized as follows. Discussing the dynamics
of interacting domain walls in Sect.~\ref{PredictionSection}, we
argue that short-range forces are not sufficient to stabilize
fluctuating ordered phases in one spatial dimension. 
In Secs.~\ref{TCPSection} and~\ref{PCPDSection} 
we revisit the triplet creation process and
the annihilation/fission process, both being examples of models
with short-range forces. As a third example a
spreading process on a diffusing background will be discussed
in Sec.~\ref{TwoSpeciesSection}. Although in this case
long-range forces lead to a coarsening process, 
the transition turns out to be continuous as well.

\section{Interacting domain walls in one dimension}
\label{PredictionSection}

\begin{figure}
\epsfxsize=85mm
\centerline{\epsffile{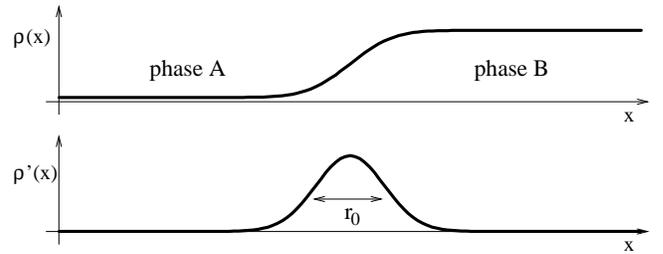}}
\caption{
\label{FIGKINK}
Schematic profile of the coarse-grained particle density $\rho(x)$
and its derivative $\rho'(x)$ at a domain wall between
phase {\bf A} and phase {\bf B} in a one-dimensional system.
}
\end{figure}

\headline{Short-range forces between domain walls}
To explain the impossibility of first-order transitions in
the processes (\ref{DickmanTomeProcess})-(\ref{AFProcess})
from a phenomenological point of view,
let us consider a hypothetical one-dimensional system with two
ordered phases {\bf A} and {\bf B}. Without loss of generality we
assume the {\bf A}-phase to be absorbing while
the {\bf B}-phase fluctuates, i.e. small {\bf A}-islands 
are spontaneously created
in the bulk of the {\bf B}-phase. Furthermore, we assume that
there is a robust mechanism which eliminates minority
islands generated by fluctuations, ensuring the stability of
phase {\bf B}. For a system with these properties, let us consider an
initial state where half of the system is in phase {\bf A}
while the other half is in phase {\bf B} (see Fig.~\ref{FIGKINK}).
Both phases are separated by a domain wall. In contrast
to the Glauber-Ising model, this domain wall is not necessarily
associated with a single broken bond, rather is may be `smeared out'
over a certain region in space. However, since both phases are
attracting,
the domain wall remains {\em localized}, i.e., it has a finite width.
Consequently, the derivative of the order parameter
profile has exponentially decreasing tails.
Obviously, the critical point of such a model corresponds to a
situation where the domain wall diffuses without bias.

In one dimension a minority islands of phase {\bf A} inside phase {\bf B}
can be considered as a pair of such domain walls. In the
Glauber-Ising model the domain walls do not interact unless they
annihilate at the same site. In our hypothetical model, however,
the domain walls do interact, giving rise to an attracting `force'
eliminating islands of the minority phase. However, since the 
domain walls have a finite width, 
we expect this `force' to have a finite range
$r_0$. More precisely, for large island sizes $r$ we expect the
strength of the interaction to decrease {\em exponentially} as
$\exp(-r/r_0)$ or even faster.

\headline{Instability of fluctuating ordered domains}
The essential problem arises precisely at this point. In one
spatial dimension an exponentially decreasing short-range force
between two domain walls is not sufficient to stabilize a
fluctuating ordered phase. In order to understand this point, let
us consider an island of size $r$ which grows by one step with
rate $1-a\exp(-r/r_0)$ and shrinks with rate $1+a\exp(-r/r_0)$. In
this toy model the parameters $a$ and $r_0$ control the strength
and the range of the force, respectively. The time evolution of
the corresponding probability distribution $P_r(t)$ is given by
the equation\footnote{In principle there may be  two different 
length scales on both sides of the domain wall.
}
\begin{equation}
\label{Evolution}
\begin{split}
\frac{\partial}{\partial t} P_r(t) =
-2P_r(t) + (1+a\,e^{-(r+1)/r_0})P_{r+1}(t) \\
+ (1-a\,e^{-(r-1)/r_0})P_{r-1}(t)
\end{split}
\end{equation}
with the boundary condition $P_0(t)=0$. Starting with a single
spin flip $P_r(0)=\delta_{r,1}$, we obtain a broadening
distribution of island sizes. Without a short range force (i.e.
$r_0=0$), Eq.~(\ref{Evolution}) reduces to the discrete diffusion
equation with a Dirichlet boundary condition at $r=0$. In this
case the asymptotic solution reads
\begin{equation}
P_r(t) \;\stackrel{t \rightarrow \infty}{\simeq}\;
C \, t^{-3/2} \, \exp(-r^2/4t)\,,
\end{equation}
where $C$ is a certain amplitude factor. In presence of a
short-range force with $r_0>0$, this asymptotic solution remains
valid -- the only change is in the prefactor $C$. Although $C$
decreases exponentially with increasing $r_0$, it is always
positive. Therefore, there is a small but finite probability that
the two domain walls become effectively unbound, producing a
macroscopic minority island with a size of the order
$\sqrt{\tau}$. Thus, the ordered phase is only marginally stable,
irrespective of the interaction range $r_0$. As a consequence, the
{\bf B}-phase eventually disintegrates, approaching the absorbing
state {\bf A}. This contradiction in our hypothetical model
demonstrates that in one spatial dimension {\it it is impossible
to stabilize fluctuating ordered phases by short-range
interactions}. It should be emphasized that these arguments are
not valid in higher dimensions where the dynamics of domain walls
may depend on the local curvature.

\headline{Properties of the continuous transition}
Since absorbing islands may reach a macroscopic size, 
our hypothetical model approaches the absorbing state.
In order to prevent the system from reaching the absorbing state,
we may now slightly modify the parameters such that {\bf B} domains
tend to grow. Since the {\bf A}-phase is absorbing, 
the resulting competition between growing {\bf B}-domains
and spontaneously created {\bf A}-islands suggest a crossover to
directed percolation (DP)~\cite{Kinzel,Hinrichsen00}, which is 
the generic universality class of continuous phase transition 
into absorbing states. For instance, as will be shown in 
Sec.~\ref{TCPSection}, the triplet creation process 
displays such a crossover to DP. 
In general, the universality class of 
the continuous transition will depend on the symmetry properties 
and the conservation laws of the process under consideration. As an example,
we will discuss the annihilation/fission (see Sec.~\ref{PCPDSection})
which crosses over to a non-DP transition.

\headline{Numerical checks}
Obviously these considerations are only valid if the ordered
states are sufficiently attracting to confine domain walls to a
finite region with a typical size $r_0$. There are several
possibilities to verify this assumption. The simplest method would
be to perform high-precision Monte Carlo simulations in order to
confirm the crossover to a continuous transition. 
Alternatively, we may check whether a
compact $B$-phase disintegrates as predicted. As a more sensitive
check, one could also measure the distribution of the sizes of
minority islands. Since spin flips occur spontaneously in the ordered
phase, we expect this probability distribution to become 
stationary in the limit $t \rightarrow \infty$.
Solving the stationary problem of
Eq.~(\ref{Evolution}), we obtain the distribution
\begin{equation}
\label{StationarySolution}
P_r = \prod_{r'=1}^{r-1} \,
\frac{1-a\,e^{-r'/r_0}}{1+a\,e^{-(r'+1)/r_0}}\,P_1\,.
\end{equation}
This distribution decays exponentially for small $r$ until it
saturates at a small but finite constant in the limit
$r \rightarrow \infty$, expressing the fact that
minority islands of all sizes will be formed.
A distribution of this form observed in a numerical simulation
can be considered as a hallmark for the presence of a 
short-range force between domain walls.

\section{The triplet creation process $3X\rightarrow 4X$, $X\rightarrow
\O$} \label{TCPSection}
%
%
%
\begin{figure}
\epsfxsize=85mm
\centerline{\epsffile{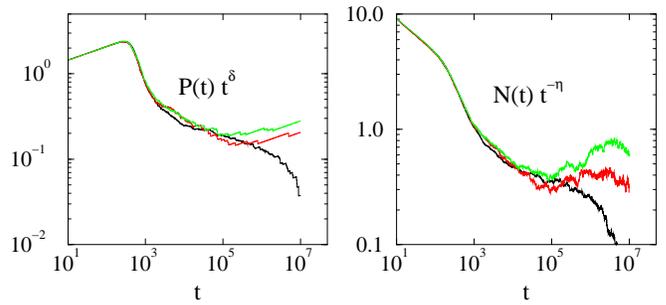}}
\vspace{2mm}
\caption{
\label{FIGPCP}
Numerical simulation of the triplet creation process.
Starting with an island of 20 particles the survival
probability $P(t)$ (left) and the mean number
of particles $N(t)$ (right) of the cluster
are averaged over $300$  independent runs.
The data points are multiplied with powers of $t$
in a way that DP critical behavior corresponds
to horizontal lines (see text).
}
\end{figure}

We are now going to apply these tests to the triplet creation
model introduced by Dickman and Tom\'e~\cite{DickmanTome91}. It
evolves by random sequential updates and is controlled by two
parameters, namely a rate for offspring production $\lambda$ and a
diffusion constant $D$. For $D<0.85$ the transition was found to
belong to the universality class of directed percolation, whereas
for $D>0.85$ a first order transition was reported. Moreover, the
phase transition line seemed to split up into two spinodal lines
$\lambda_\pm(D)$. In the following we show that the transition
crosses over to DP after very long time. We restrict the analysis
to the case $D=0.9$, where Dickman and Tom\'e reported the
stability limits $\lambda_-=10.12(1)$ and $\lambda_+=10.30(2)$.

As a first test, we perform Monte Carlo simulations starting with
a single island of particles at the
origin~\cite{GrassbergerTorre79}. It turns out that it is
necessary to start with an island of several particles since
otherwise the survival probability of the cluster is extremely
small. At criticality the survival probability $P(t)$ and the
average number of particles $N(t)$ are expected to obey asymptotic
power laws of the form
\begin{equation}
P(t) \sim t^{-\delta}\,, \qquad N(t) \sim t^\eta
\end{equation}
with $\delta=1/2, \eta=0$ for CDP (indicating a first-order
transition) and $\delta \simeq 0.159, \eta \simeq 0.313$ in the
case of a DP transition, respectively. The results of the
simulations are shown in Fig.~\ref{FIGPCP}. Obviously, there are
three different temporal regimes. In the first $100$ time steps,
the island survives with certainty due to the large initial size
of $20$ sites, followed by a rapid decrease of $P(t)$ faster than
$1/\sqrt{t}$. This part of the temporal evolution is expected to
be non-universal. Then the model enters a second regime extending
over two decades from $10^3$ to $10^5$ time steps, where $P(t)$
decays as $1/\sqrt{t}$ and $N(t)$ stays almost constant. This is
the time window where the model behaves essentially in the same
way as a zero-temperature Glauber-Ising model so that the
transition appears to be discontinuous. In the following two
decades from $10^5$ to $10^7$ time steps, however, we observe a
slow crossover to DP exponents. Our estimate for the critical
point $\lambda=10.145(10)$ lies between the stability limits
$\lambda_\pm$ reported in~\cite{DickmanTome91}. With the computer
technology ten years ago, Dickman and Tom\'e could only go up to
$22000$ time steps so that they were unable to observe the
crossover to DP. A similar crossover phenomenon from an initial
transient over CDP to DP was recently observed in certain models
for flowing sand on an inclined plane~\cite{Sand}.

\begin{figure}
\epsfxsize=70mm \centerline{\epsffile{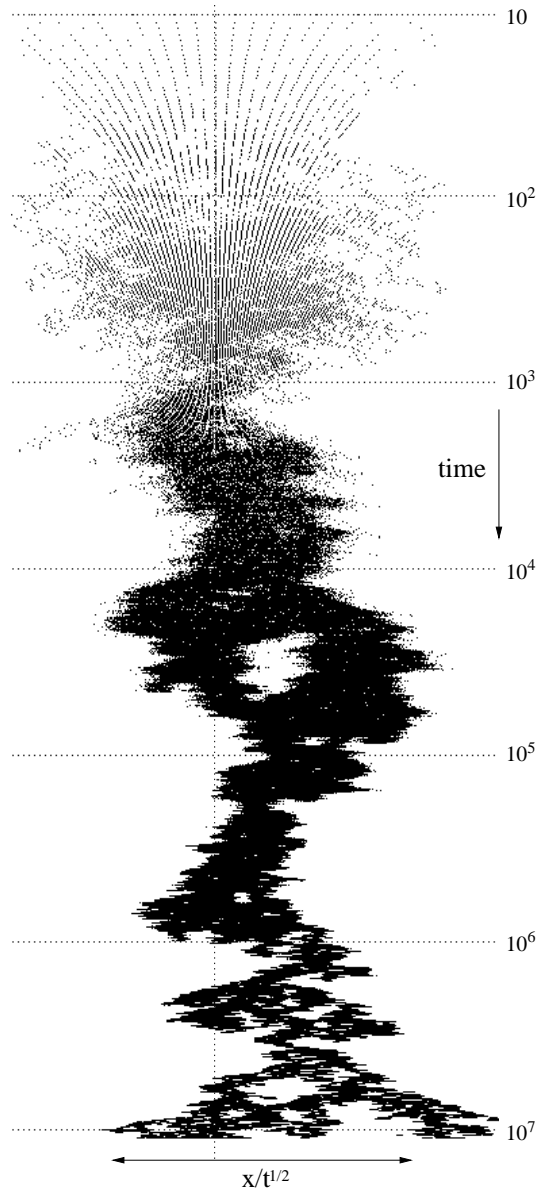}} \vspace{2mm}
\caption{ \label{FIGDEMO1} The triplet creation process
$3X\rightarrow 4X$, $X\rightarrow \O$ at the critical point. The
figure shows the temporal evolution over seven decades starting
with a compact island of 20 particles in the center. The graphs
shows the rescaled position $x/\sqrt{t}$ of the particles as a
function of $\log(t)$ measured in a single run. Up to $10^4$ time
steps the cluster has the form of a compact diffusing cloud of
particles with very small islands of unoccupied sites generated by
fluctuations. The first macroscopic minority island emerges after
$2\cdot10^4$ time steps, indicating the beginning of the crossover
to DP which extends up to $10^6$ steps. Only in the last decade,
where the thickness of active branches is small compared to the
lateral cluster size, we observe the typical patterns of a
directed percolation process. }
\end{figure}

As a second check, we demonstrate that the fluctuating phase
disintegrates, generating the typical patterns of DP clusters
after very long time. To this end we introduce a novel type of
space-time plots which can be used to visualize the scaling
properties of critical clusters in systems with absorbing states
(see Fig.~\ref{FIGDEMO1}). Starting with a localized seed (an
island of 20 particles) at the origin and simulating the process
up to $10^7$ time steps, the rescaled position of the particles
$x/t^{1/z}$ is plotted against $\log_{10} t$, where
$z=\nu_\parallel/\nu_\perp$ is the dynamic exponent of the process
under consideration. Here we choose $z=2$ since the domain walls
are expected to diffuse. By rescaling the spatial coordinate $x$,
the cluster evolves within a strip of finite width. Unlike linear
space-time plots the scale-invariant representation of a cluster
allows one to survey more than six decades of the temporal
evolution. As can be seen, the figure is consistent with the Monte
Carlo simulation results. After an initial transient the cluster
appears to be compact, behaving essentially in the same way as in a
Glauber-Ising model at zero temperature. The boundaries, which are
smeared out by diffusion, become sharper as time proceeds. The
first minority island with a macroscopic size appears after
$2\cdot 10^4$ time steps, where the crossover to DP begins.
However, the typical patterns of a DP cluster can only be seen
after $10^6$ time steps.

\begin{figure}
\epsfxsize=75mm
\centerline{\epsffile{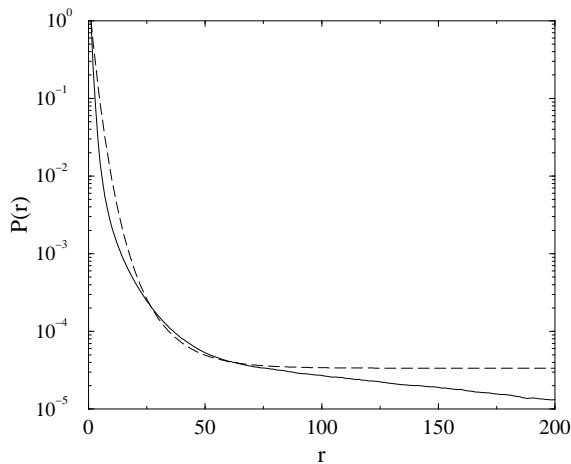}}
\caption{
\label{FIGISLANDS}
Distribution of the sizes of inactive islands in the
critical triplet creation model (solid line), compared
to a fit of Eq.~(\ref{StationarySolution}) (dashed line).
}
\end{figure}

As a third check, we measure the stationary distribution $P_r$ of
island sizes in order to verify the predictions of the toy model
in Sect.~\ref{PredictionSection}. Starting with a fully occupied
lattice, the sizes of inactive islands are measured in a critical
system of 10000 sites after an `equilibration' time of $10^6$ time
steps. The result is shown in Fig.~\ref{FIGISLANDS} as a solid
line. As can be seen, the distribution $P_r$ decays exponentially
or even faster by almost five decades until it crosses over to a
slowly decreasing function. Fitting the parameters $a$ and $r_0$,
this curve may be compared with the stationary solution of the toy
model in Eq.~(\ref{StationarySolution}). Although the curves do
not coincide, they show the same qualitative behavior. This
supports the point of view that domain walls in the
1+1-dimensional triplet creation process attract one another by a
short-range force so that the fluctuating active phase cannot be
stabilized.

%
%

\section{The annihilation/fission process $2X\rightarrow 3X$,
$2X\rightarrow \O$} \label{PCPDSection}

Some time ago Howard and T{\"a}uber introduced the so-called
annihilation/fission process which exhibits an unconventional
nonequilibrium phase transition~\cite{HowardTauber97}. Their
original motivation was to consider a reaction-diffusion process
which allows one to interpolate between `real' and `imaginary'
noise in the corresponding Langevin equation. Performing a
field-theoretic renormalization group analysis they predicted
non-DP critical behavior at the transition.

More recently, Carlon {\em et al.}~\cite{CHS99} studied a lattice
version of the model which is defined  by the dynamic rules
\begin{equation}
\begin{split}
AA\text{\O},\,\text{\O} AA \rightarrow AAA  \qquad \text{with rate}
\, & (1-p)(1-d)/2 \\
AA \rightarrow \text{\O\O} \qquad  \text{with rate}\,  &
p(1-d) \\
A\text{\O} \leftrightarrow \text{\O}A \qquad \text{with
rate}\,  & D \ .
\end{split}
\label{DynamicRules}
\end{equation}
Performing a density matrix renormalization group analysis they
arrived at the conclusions that for low values of the diffusion
constant $D$ the transition is continuous while it becomes first
order for higher values of $D$ above a certain tricritical point.
The paper by Carlon {\it et al.} released a debate concerning the
universality class of the continuous phase transition. Because of a numerical
coincidence of two out of four critical exponents, the authors 
concluded that the transition should belong to the
parity-conserving universality
class~\cite{TakayasuTretyakov92,ALR93,CardyTauber96,Menyhard94,KimPark94,Hinrichsen97}.
This result was questioned in Ref.~\cite{Hinrichsen00b} since
there is neither a parity conservation law nor a $Z_2$-symmetry in
the AF process. Moreover it was pointed out that a first order
transition might not exist in one spatial dimension. Increasing
the diffusion constant, continuously varying critical exponents were
reported in~\cite{Odor00}. For large values of $D$ these exponents seemed
to approach certain values which can also observed in cyclically
coupled DP and annihilation processes~\cite{Hinrichsen00c}. So far
there is no numerical evidence for a discontinuous transition.

\begin{figure}
\epsfxsize=65mm
\centerline{\epsffile{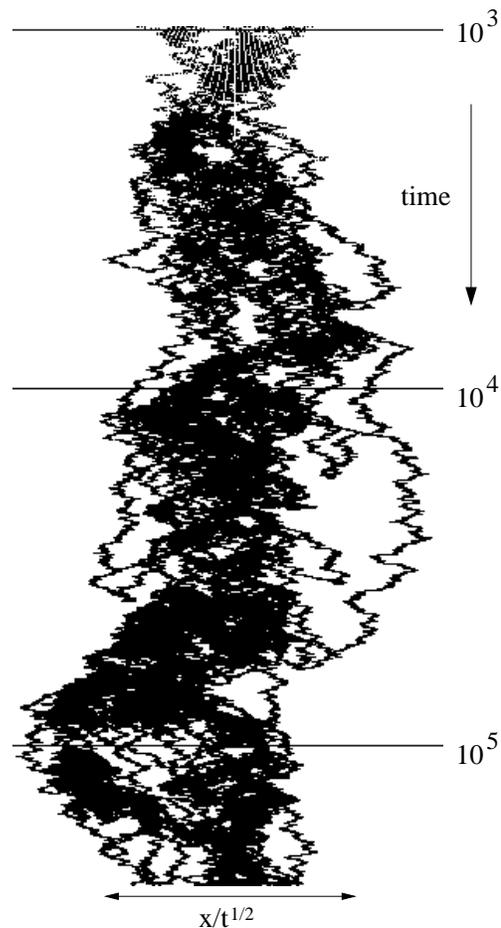}}
\caption{
\label{FIGAF}
Scale-invariant plot of a space-time history
of the annihilation/fission process for $D=0.9$ at the
critical point $p_c=0.233(2)$. As can be seen,
the high density phase disintegrates, creating
macroscopic islands of the minority phase in the bulk.
}
\end{figure}

It should be emphasized that the AF process has two non-symmetric
absorbing states, namely the empty lattice and the state with a
single diffusing particle. The unusual critical behavior is
related to the fact that solitary particles may diffuse over large
distances before they meet another particle in order to annihilate
or to release a new avalanche. Therefore, the absorbing state may
be `less attracting' than in the triplet creation model. Because
of these special properties, it is not immediately clear whether
the assumptions of Sect.~\ref{PredictionSection} apply. For
example, solitary diffusing particles could lead to an effective
long-range interaction between domain walls. However, as
demonstrated in Fig.~\ref{FIGAF}, even for high values of the
diffusion rate a compact island quickly disintegrates, generating
nontrivial patterns of active patches and solitary particles.
This supports the claim by \'Odor~\cite{Odor00} that there
is no first-order phase transition in the AF process for any
$0<D<1$. The question whether the observed continuously varying
critical exponents are related to asymptotically well-defined
properties or simply to crossover effects between an an effective 
first-order behavior on short scales and an asymptotic
continuous transition remains open.

%
%

\section{Spreading process on a diffusing background}
\label{TwoSpeciesSection}
%
%
Let us finally investigate the following two-species particle
process with random-sequential updates:
\begin{eqnarray}
\label{XYProcess}
& X+Y \rightarrow Y+Y         & \quad \text{with rate} \quad  \alpha \nonumber \\%
& Y \rightarrow X             & \quad \text{with rate} \quad  \beta \\%
& X+\O \leftrightarrow \O+X   & \quad \text{with rate} \quad  D_X \nonumber\\%
& Y+\O \leftrightarrow \O+Y   & \quad \text{with rate} \quad  D_Y
\nonumber
\end{eqnarray}
This reaction diffusion scheme may be interpreted as a spreading
process of $Y$ particles on a diffusing background of $X$
particles. In this process the total number of particles is
conserved. Therefore, the spreading properties are controlled by
the density of particles in the initial state
$\rho=\rho_X+\rho_Y$, while the density $\rho_Y$ plays the role of
an order parameter. For low values of $\rho$ the particles are
sparsely distributed so that the system quickly evolves into an
`absorbing' state without $Y$ particles. For high values of $\rho$
the spreading process $X+Y\rightarrow 2Y$ dominates, leading to a
stationary density $\rho_Y>0$ in a sufficiently large system. Both
phases are separated by a nonequilibrium phase transition at a
certain critical density $\rho_c$ which depends on the parameters
$\alpha,\beta$ and the diffusion rates $D_X$ and $D_Y$.

The critical behavior of the model depends significantly on the
ratio of the diffusion rates $D_X$ and $D_Y$. If both rates are
equal, the particles diffuse in the same way as in an ordinary
exclusion process, whereas the labels $X$ and $Y$ 
can be considered as `colors' with separate dynamic rules.
This special case has been considered in
Ref.~\cite{EqualDiffusionRates}, 
where a continuous phase transition was found.
Recently, van Wijland {\it et al.} were able to confirm these results
by a field-theoretic analysis~\cite{WilsonRenormalization}.
Moreover, they investigated the general case $D_X \neq D_Y$. For
$D_X < D_Y$ they predicted a continuous phase transition with
non-DP critical exponents, while for $D_X>D_Y$ it turned out to
be impossible
to perform a controlled field-theoretic analysis. For this case
van Wijland {\it et al.} conjectured that the transition might be
discontinuous.

\begin{figure}
\epsfxsize=85mm \centerline{\epsffile{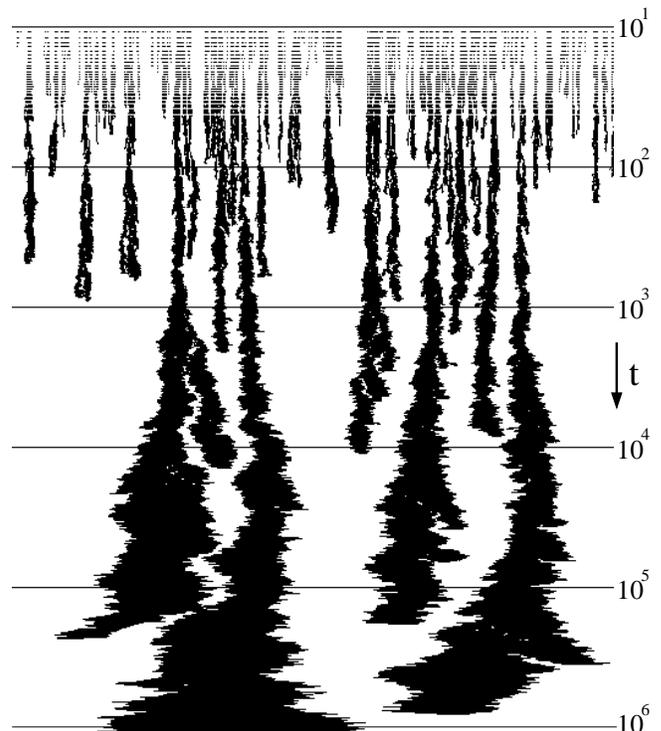}}%
\caption{ \label{FIGCHANNELS} 
Semi-logarithmic plot of the spreading process~(\ref{XYProcess})
with $1000$ sites
for $D=0.9$ in the active phase $\rho-\rho_c$. The figure shows
the dynamics of the $Y$ particles, while the diffusing background
of $X$ particles is not visible. As can be seen, islands of $Y$
particles coarsen until a single island is left.
}
\end{figure}

Obviously, the arguments used in
Sect.~\ref{PredictionSection} cannot be applied to this model
since the particle conservation law introduces 
effective long-range interactions.
Thus, the model may well be a candidate for
a first-order transition even in one
spatial dimension. To investigate this question in more detail, we
analyzed the two-species model numerically. Choosing $\alpha=0.2$,
$\beta=1$ and $D_Y=1-D_X$, the process (\ref{XYProcess}) is
controlled by two parameters $D_X=D$ and $\rho$. The $Y$ particles
may be considered as the active sites of a spreading process
running on top of a diffusing background of $X$ particles. For $D
\leq 0.5$ the transition is characterized by fractal clusters
reminding of DP, but apparently in a different universality class.
For $D>0.5$, however, there is indeed a clear signature of
first-order behavior. As shown in the space-time plot in
Fig.~\ref{FIGCHANNELS}, the $Y$ particles are confined to certain
vertical `channels'. As time proceeds, small channels disappear
while larger ones tend to grow, leading to a coarsening process.

The mechanism behind this coarsening process can be understood as
follows. Because of $D_Y<D_X$, a high concentration of $Y$
particles reduces the effective diffusion rate. Since a locally
reduced diffusion rate leads to particle clogging, the local
density of particles increases. Thus there is positive feedback
amplifying the particle density (and hence the effective spreading
rate) in regions where the spreading process is active. Because of
this instability, the $Y$ particles tend to form compact domains
with a high particle density. In $d\geq 2$ spatial dimensions,
such an instability may lead to a first-order transition. In
one-dimensional systems, however, the nonlinear amplification mechanism
alone is not sufficient, especially when the high-density phase
fluctuates. 
Obviously, the essential mechanism behind the coarsening process
is the conservation of the total number of particles. For example,
let us assume that a high-density domain has been created by
fluctuations. Because of the instability, the spreading process is
supercritical inside the domain. Particle conservation ensures
that such an island cannot split up into two separate islands
since there is no way to reduce the density inside.

The temporal evolution of the coarsening process depends on the
dynamics of domain walls between regions with low and high
particle density. Obviously, high-density islands tend to grow
until the particle density of their environment decreases below a
certain threshold. This threshold, however, depends not only on
the initial particle density $\rho$ but also on the total size of
all active islands. In other words, active islands continue to
grow until they bind so many particles that the average density of
particles in their environment is no longer sufficient to sustain
the growth process. Moreover, large islands seem to be more stable
than small ones, leading to a slow coarsening process. Thus,
finite systems evolve into a state where only one high-density
domain survives.

\begin{figure}
\epsfxsize=80mm \centerline{\epsffile{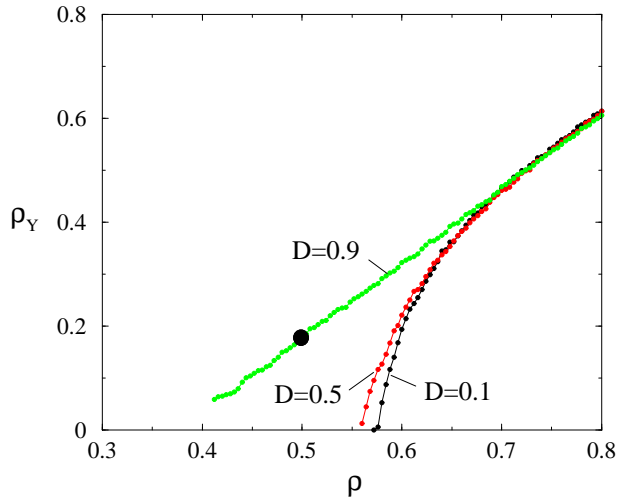}}%
\caption{ \label{FIGORDERPAR} 
Spreading process on a diffusing background. Stationary
density of $Y$-particles as a function of the total
particle density $\rho$ for various values of the diffusion
rate $D=D_x=1-D_Y$. The bold dot marks the location where
the space-time plot in Fig.~\ref{FIGCHANNELS} has been generated. }
\end{figure}

The conservation law has another surprising consequence: Although
the system coarsens, the transition is still continuous. In fact,
plotting the asymptotic stationary value of $\rho_B$ against
$\rho-\rho_c$, there is no discontinuity (see
Fig.~\ref{FIGORDERPAR}). This can be explained by observing that
the size of the surviving active domain grows almost linearly with
$\rho-\rho_c$. Therefore, the reaction-diffusion (\ref{XYProcess})
provides an interesting example of a system with an instability
which exhibits a second-order phase transition in one dimension,
although the arguments Sect.~\ref{PredictionSection} cannot be
applied.

\section{Conclusions}

First-order transitions in 1+1-dimensional nonequilibrium models
with fluctuating ordered phases require a robust mechanism
which eliminates islands of the minority phase generated by
fluctuations. In many cases 
this mechanism relies on special properties such as
the interplay of several species of particles, competing currents,
unconventional conservation laws, or special boundary conditions.
In this paper we have presented physical
arguments why first-order phase transitions are impossible
in simple two-state models without such attributes
since short-range interactions between domain walls cannot
stabilize a fluctuating ordered phase.
As an example we have revisited the triplet-creation model
which was believed to display a first-order phase transition.
However, the first-order behavior is a transient phenomenon
and eventually crosses over to a continuous phase transition,
supporting the DP conjecture by Jannsen and
Grassberger~\cite{Janssen81,Grassberger82}.
Similarly, the annihilation/fission process crosses over to 
a second-order phase transition. Thus, first-order phase transitions
in one spatial dimension are subtle and
require a much more robust mechanism for the 
elimination of minority islands.


\end{document}